\documentclass[aps,preprint]{revtex4-1}
\usepackage{graphicx,psfrag}
\usepackage{graphicx} 
\usepackage{bm}
\usepackage{booktabs}
\usepackage{amsmath,amsthm,amssymb}
\usepackage{gt12}

\usepackage{color}
\begin{document}

\title{
Active electromagnetic metamaterial based on spin torque oscillators
}
\author{  Gen Tatara$^1$,$^2$, Hiroaki T. Ueda$^1$,  Katsuhisa Taguchi$^1$, Yuta Sasaki$^1$, Miyuki Nishijima$^1$, 
Akihito Takeuchi$^2$}  
\affiliation{$^1$ Graduate School of Science and Engineering, Tokyo Metropolitan University, Hachioji, Tokyo 192-0397 Japan \\
$^2$ RIKEN Advanced Science Institute, 
2-1 Hirosawa, Wako, Saitama, 351-0198 Japan\\
$^3$ Department of Applied Physics, The University of Tokyo, 7-3-1 Hongo, Tokyo 113-8656 Japan 
}
\date{\today}
\begin {abstract} 
We propose theoretically an active material for electromagnetic radiation with frequency of GHz by use of spin-torque oscillators.
The origin of the amplification is the energy supplied to the magnetization by the injected current.
We show that close to a resonance with current-driven
magnetization, the imaginary part of magnetic permeability becomes indeed
negative for either of the two circular polarizations, resulting in negative
imaginary part of refractive index.
Besides, the real part of the refractive index is also manipulated by the current.
Our system thus realizes an active filter to obtain circular polarized radiation and/or an electromagnetic metamaterial having negative refractive index, both controlled electrically.

\end{abstract}
%

\newcommand{\Atil}{\tilde{A}}
\newcommand{\Btil}{\tilde{B}}
\newcommand{\mutil}{\tilde{\mu}}
\newcommand{\hv}{{\bm h}}
\newcommand{\Hext}{H_{\rm ext}}
\newcommand{\Hvext}{\Hv_{\rm ext}}
\newcommand{\hext}{h}
\newcommand{\Hvem}{\Hv_{\rm em}}
\newcommand{\hvem}{\hv_{\rm em}}
\newcommand{\jmeta}{j_{\rm meta}}
\newcommand{\omegac}{\Omega_{\rm c}}
\newcommand{\omegaM}{\Omega_{M}}
\newcommand{\omegaP}{\omega_{\rm p}}
\newcommand{\Mvf}{\Mv_{\rm f}}

\maketitle

\section{Introduction}

Electric control of material properties is highly important in technologies.
For instance, manipulation of magnetic structure by electric current is promising for ultra high density non-volatile memories  \cite{Parkin08} and logics \cite{Allwood05}, and control of response to electromagnetic radiation is useful for sensing, imaging and in communication devices \cite{Soukoulis11}.
Here we propose a mechanism of electrically-driven transparency and amplification in ferromagnetic metals in the GHz range.
The system is a bilayer of thin ferromagnets under a magnetic field and DC current.
When current is applied to the bilayer, precession of magnetization of the free layer is induced \cite{Slonczewski96,Berger96}, resulting in a spin-torque oscillator \cite{Kiselev03,Deac08}.
The spin-transfer torque induced by the applied current acts as a negative damping \cite{Berger96} and then spin-torque oscillator becomes an electromagnetic active media close to the resonance frequency.
This means that a bilayer of ferromagnetic metals, which is a perfect reflector of microwaves, becomes transparent when a current is applied.
This transparency is a result of active nature of the system, and thus the transmitted wave is amplified.
The system exhibits in addition a significant effect of negative real part of the refractive index, i.e., electromagnetic left-handed metamaterial \cite{Pendry96}. 
The present mechanism therefore can be applied to current-induced switch, amplifier, polarizer and beamsplitter for GHz waves.

Negative index of refraction is one of the most counter-intuitive and
fascinating phenomena in the physics of electromagnetic field, where a beam of
radiation incident on an interface between two materials is refracted towards a
wrong direction \cite{Smith04}.
Such a possibility was theoretically proposed by Veselago many years ago \cite{Veselago68}.
He discussed that the real part of index of refraction becomes negative if the real parts of
both permittivity $\epsilon$ and magnetic permeability $\mu$ are negative, and
also that negative $\epsilon$ and $\mu$ arise when a resonance occurs in both
electric and magnetic properties.
In naturally occurring materials, however, negative index of refraction is not common.
This is because the electric properties of most metallic materials are governed by frequencies higher than THz (meV-eV order as the energy scale), while frequencies for magnetic properties are lower than 100GHz, and thus electric and magnetic resonances do not usually occur simultaneously.
One way to overcome this difficulty is 
 to construct an artificial material, a metamaterial, to realize a system where the plasma frequency, of the order of  2400THz (corresponding to 10 eV)
in metals, is shifted to lower frequencies and/or the magnetic
resonance occurs at higher frequencies.
A possibility to lower electric resonance was given by Pendry by use of split
ring resonators, where induced currents in rings result in magnetic resonance
\cite{Pendry96}, and experimental realization of negative index of refraction
was done in the microwave regime (frequency of 4.8GHz) by combining split ring
resonators of millimeter size with continuous wires \cite{Smith00}.
Later, by use of smaller resonators of 20$\mu$m size, the frequency was
increased to THz regime \cite{Moser05}.
Besides split-ring resonators, multilayer structure with arrays of holes was used to realize negative index of refraction in near-infrared regime (around 150THz) \cite{Zhang_meta05}.
Very recently, it was demonstrated that negative refractive index in the GHz range is realized in a thin film of natural ferromagnetic metal \cite{Pimenov07,Engelbrecht11}.

For applications, realization of low loss material is of crucial importance.
Combining negative refraction medium with gain medium (active medium) formed by
two-level emitters \cite{Bergman03} is one possibility \cite{Sarychev07}.
In this paper, we propose an active medium in the GHz regime based on a
current-driven magnetization of small metallic ferromagnets.
The system can indeed acquire a negative real part of the refractive index at the same
time, realizing an active electromagnetic metamaterial with negative refractive
index.
(In this paper, we call systems with negative refractive index the
electromagnetic metamaterials.)

In the case of ferromagnetic resonance driven by an external magnetic field,
damping is represented by the Gilbert damping constant \cite{Chikazumi97},
$\alpha$, which is positive definite.
In other words, a static external magnetic field cannot transfer energy to the precessing
magnetization, since the motion induced by its torque is always perpendicular to
the applied field.
This situation changes significantly when magnetization dynamics is driven by
applying an electric current.
As pointed out by Slonczewski and Berger \cite{Slonczewski96,Berger96}, when an
electric current is injected in a layered structure of two ferromagnets, each
ferromagnet exerts a torque on the magnetization of the other ferromagnet.
There occur two types of torques; a spin-transfer torque
\cite{Slonczewski96,Berger96} and a field-like torque 
\cite{Thiaville05,Zhang04,TKS_PR08,Sankey08}.
The field-like torque acts the same way as a magnetic field, while spin-transfer
torque is unique for current-induced case.
Choosing the fixed layer magnetization to be along the precession axis, this torque becomes parallel to the damping torque, and thus can reduce the effective
damping torque even to a negative regime \cite{Berger96}.
When the effective damping  becomes negative, magnetization starts spontaneous
precession, resulting in spin-torque oscillators \cite{Kiselev03}.
If an electromagnetic wave with frequency close to the precession frequency is
injected to such system, the imaginary part of magnetic permeability becomes
negative.
The current-driven magnetization thus works as a gain medium, if the
imaginary part is sufficiently large in the negative direction to compensate the
dissipation in the electric response.
We will discuss the real and imaginary parts of the refractive index by taking
account of the permittivity and present a phase diagram in the plane of
frequency and applied current.
We will demonstrate that a gain medium is indeed realized for circularly polarized wave
if we tune the frequency close to the resonance by use of an external magnetic field.
The condition for active material turns out, however, to be rather hard to realize; 
The current density needs to be tuned within the accuracy of 0.1\% in the worst case.
This condition is relaxed by designing a metamaterial to lower the plasma frequency like done in Ref. \cite{Pendry96}.

In the next section, we describe a spin-torque oscillator and study its response to an incident electromagnetic wave. In Sec. \ref{SEC:epsilon}, we briefly discuss the permittivity, and discuss refractive index in Sec. \ref{SEC:n}.

\section{Magnetic permeability}

\begin{figure}\centering
\includegraphics[width=0.5\hsize]{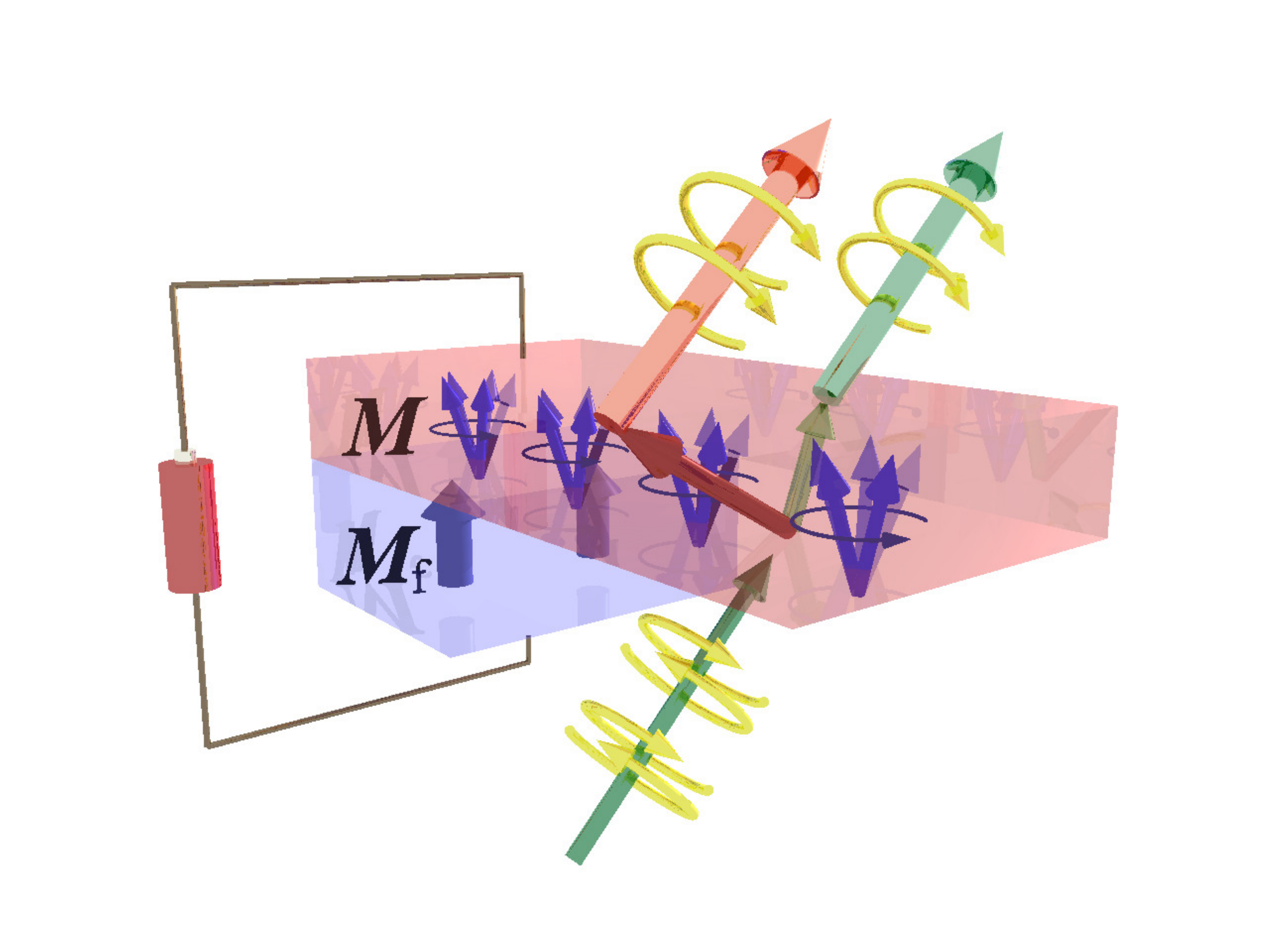}
\includegraphics[width=0.4\hsize]{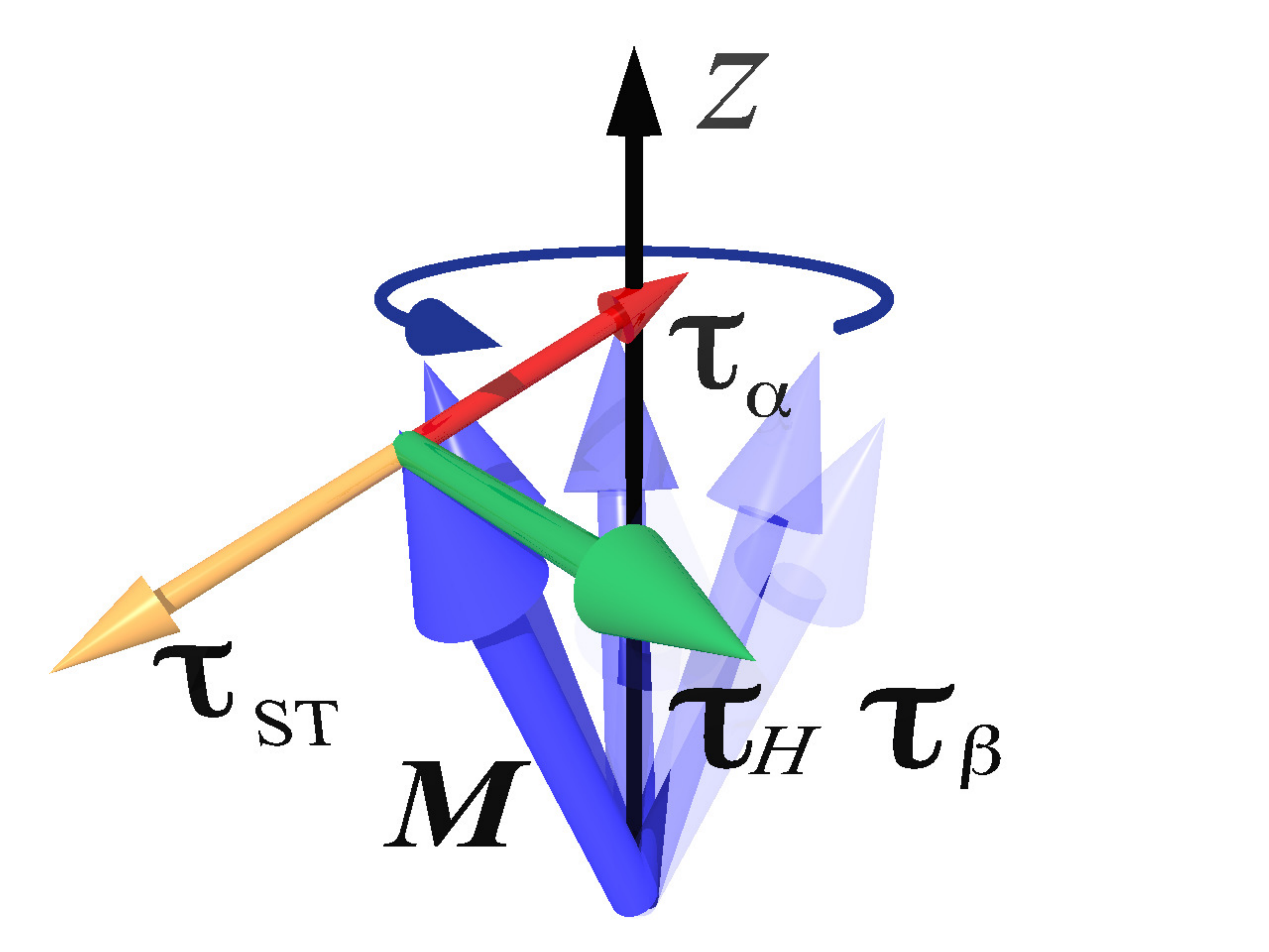}
\caption{
(a) Schematic figure of the device consisting of two ferromagnetic layers. 
Lower layer is a ferromagnet with a magnetization $\Mvf$ fixed in the $z$
direction, and upper layer is a thin ferromagnet whose magnetization $\Mv$ can
precess when microwave and/or current are applied. 
When the condition for the active media is satisfied for right-handed
circular polarization, our system works as a active filter to amplify that
particular polarization (outgoing green arrow).
If the system is in the active meta region, i.e., if both real and imaginary parts of refractive index are negative, the transmitted wave is circularly
polarized and refracted towards a wrong direction (red arrow).
The system in this regime works as an active beam splitter.
(b)
A schematic figure of torques acting on a local spin with large $z$ component
and precessing in the $xy$ plane.
The external field $\Hvext$ and fixed layer spin $\Mvf$ is along $z$ axis, and
the magnetic field of the incident microwave, $\Hvem$, is in the $xy$ plane. 
The torque due to $\Hvext$, $\torque_{\rm H}(\equiv -\muz\gamma \Mv\times \Hvext)$, and $\beta$ torque,
$\torque_\beta$, point towards the tangent to the precession.
The damping torque, $\torque_\alpha=\frac{\alpha}{M}\Mv\times\dot{\Mv}$, tends
to tilt the precession towards the equilibrium direction along $z$ axis.
Spin-transfer torque, $\torque_{\rm ST}$, acts opposite to the damping torque if
$\jtil<0$ for a right-handed polarized wave, resulting in a negative damping.
\label{FIGmeta_system}}
\end{figure}

Our system consists of two metallic ferromagnetic layers separated by a thin
insulator as is a standard setting for current-driven magnetization reversal
(Fig. \ref{FIGmeta_system}(a).
One ferromagnet is a free layer, whose magnetization (local spins) precesses
when current or external field is applied, while the other, fixed layer, has a
fixed magnetization.
When electric current is injected perpendicular to the junction, the free layer
feels two torques, a spin-transfer torque, $\torque_{\rm ST}$ \cite{Slonczewski96,Berger96}, and a
field-like torque (perpendicular torque), $\torque_{\beta}$ \cite{Zhang04,Thiaville05,Deac08}.
Denoting magnetizations of free and fixed layers by $\Mv$ and $\Mvf$, respectively,
these torques are
\begin{align}
\torque_{\rm ST} = \frac{ja^2}{eS^2} P \Mv\times(\Mv\times \Mvf) ,
& \;\;\;
\torque_{\beta} = \frac{ja^2}{eS} \beta \Mv\times \Mvf,
\end{align}
where $j$ is the applied current density, $P$ is a numerical constant
proportional to the spin splitting of the conduction electron, $\beta$ is a
constant representing the strength of spin relaxation, $M\equiv|\Mv|$ and $a$ is
lattice constant.

Our first aim is to study the behavior of magnetic permeability when an
electromagnetic wave is injected into the free layer.
We thus include the magnetic field of the electromagnetic wave, $\Hvem$.
The free layer is thinner than the penetration length, which is $\mu$m scale for
a GHz case.
For generality, we also apply an external magnetic field, $\Hvext$.
In $\Hvext$, we include the effect of magnetic anisotropy field of the system.
The equation of motion for $\Mv$ then reads \cite{TKS_PR08}
\begin{align}
\dot{\Mv} &= -\muz\gamma \Mv\times (\Hvext+\Hvem)
             +\frac{\alpha}{M}\Mv\times \dot{\Mv}
 +\jtil\lt[\frac{P}{M} \Mv\times(\Mv\times\Mvf)+\beta(\Mv\times\Mvf)\rt],
\label{LLG}
\end{align}
where $\jtil\equiv \frac{a^2}{eS^2}j $,  $\muz$ is the permeability in the
vacuum, $\gamma(=\frac{ge}{2m}>0)$ is gyromagnetic ratio ($g$ is g-factor and electron charge is $-e<0$) 
and $\alpha$ is Gilbert damping constant representing relaxation of magnetization.

We choose the external magnetic field to be along $z$ axis and fixed layer spin as $\Mvf=M(0,0,1)$.
In this case, the field-like torque points the same direction as
the torque from the external field, and the dominant part of the spin-transfer
torque is parallel or antiparallel to the damping torque, as seen from Eq.
(\ref{LLG}) as follows;
$\Mv\times(\Mv\times\Mvf)\propto \Mv\times(\Mv\times\Hvext)\propto
\Mv\times \dot{\Mv}+O(\alpha,\beta,\jtil,\Hem)$.
The direction of the free magnetization, $\Mv$, is
along $z$ direction when the electromagnetic field is absent.

The electromagnetic field is injected parallel to the $z$ axis, namely, $\Hvem$
has only $x$ and $y$ components.
When $\Hvem$ is applied, the spin thus has a large $z$ component and only small
components $M_x$ and $M_y$ in the $xy$ plane, i.e., $\Mv\simeq(M_x,M_y,M)$.
We treat $M_x$ and $M_y$ to the linear order. 
For this geometry, use of $\Hv$ instead of $\Bv$ is convenient, since the components of $\Hv$ parallel to the interface are equal for both sides of the interface.
We treat the magnetic field of the injected electromagnetic field by a plane
wave, namely, $\Hvem\propto e^{i(\kv\cdot\rv-\omega t)}$, where $\kv$ is wave
vector and $\omega$ is angular frequency.
Below, we neglect the position dependence, considering a small device compared
with the wavelength (about 3 mm for a 100 GHz field).

We consider spin dynamics with angular frequency of $\omega$, i.e., 
$M_x,M_y \propto e^{-i\omega t}$.
Defining a two-component vector describing spin fluctuation, 
$\mv\equiv(M_x,M_y)$, and 
$\hvem\equiv({\Hem}_{,x}, {\Hem}_{,y})$, the linearized equation reads
($\sigma_y$ is a Pauli matrix)
(The derivative of the $z$ component, $\dot{M_z}$, is second order in
$M_x$, $M_y$ and $\Hem$.)
\begin{align} 
(-\omega-iP\jtil+(\muz\gamma \Hext+\beta \jtil+i\omega\alpha)\sigma_y)\mv=\muz\gamma M\sigma_y \hvem.
\end{align}
The magnetic permeability of the ferromagnet, $\mu$, defined by 
$\mu\hvem=\muz\hvem+\Mv$
($\muz$ is the permeability in the
vacuum) is thus obtained as  
(in a $2\times2$ matrix notation in the $xy$ plane)
\begin{align}
\mu 
&= \lt[ 1 +   
  \omegaM
 \frac{1}{\hext +\beta \jtil- i\omega\alpha -(\omega+iP\jtil)\sigma_y} \rt],
\label{muresult1}
\end{align}
where $\hext\equiv \muz \gamma \Hext$ and 
 $\omegaM\equiv \frac{\hbar\gamma^2 S}{a^3}\muz$ is an angular frequency due to
saturation magnetization.

\begin{figure}\centering
\begin{minipage}{0.45\hsize}
\includegraphics[width=0.95\hsize]{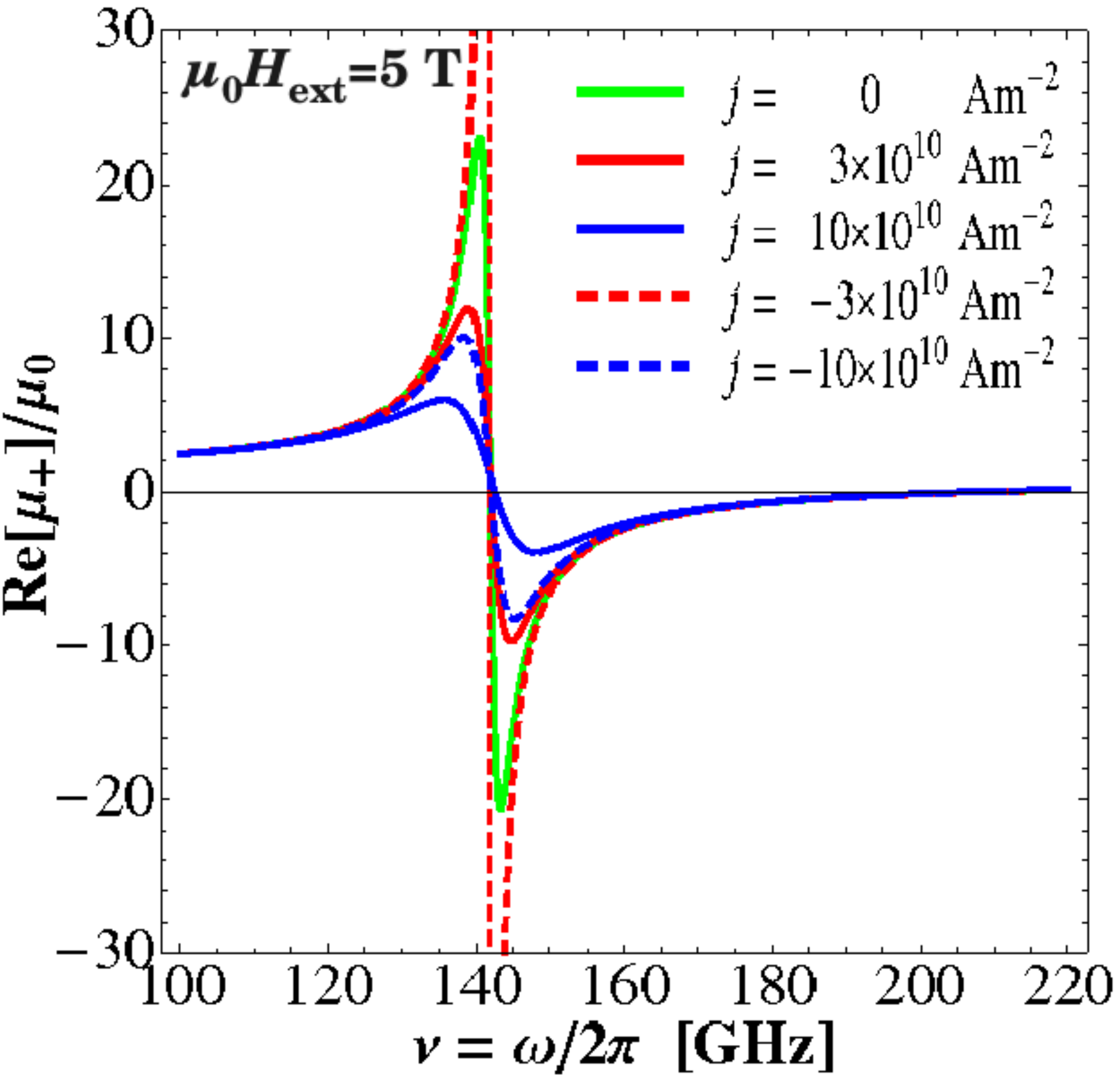}
\end{minipage}
\begin{minipage}{0.45\hsize}
\includegraphics[width=0.95\hsize]{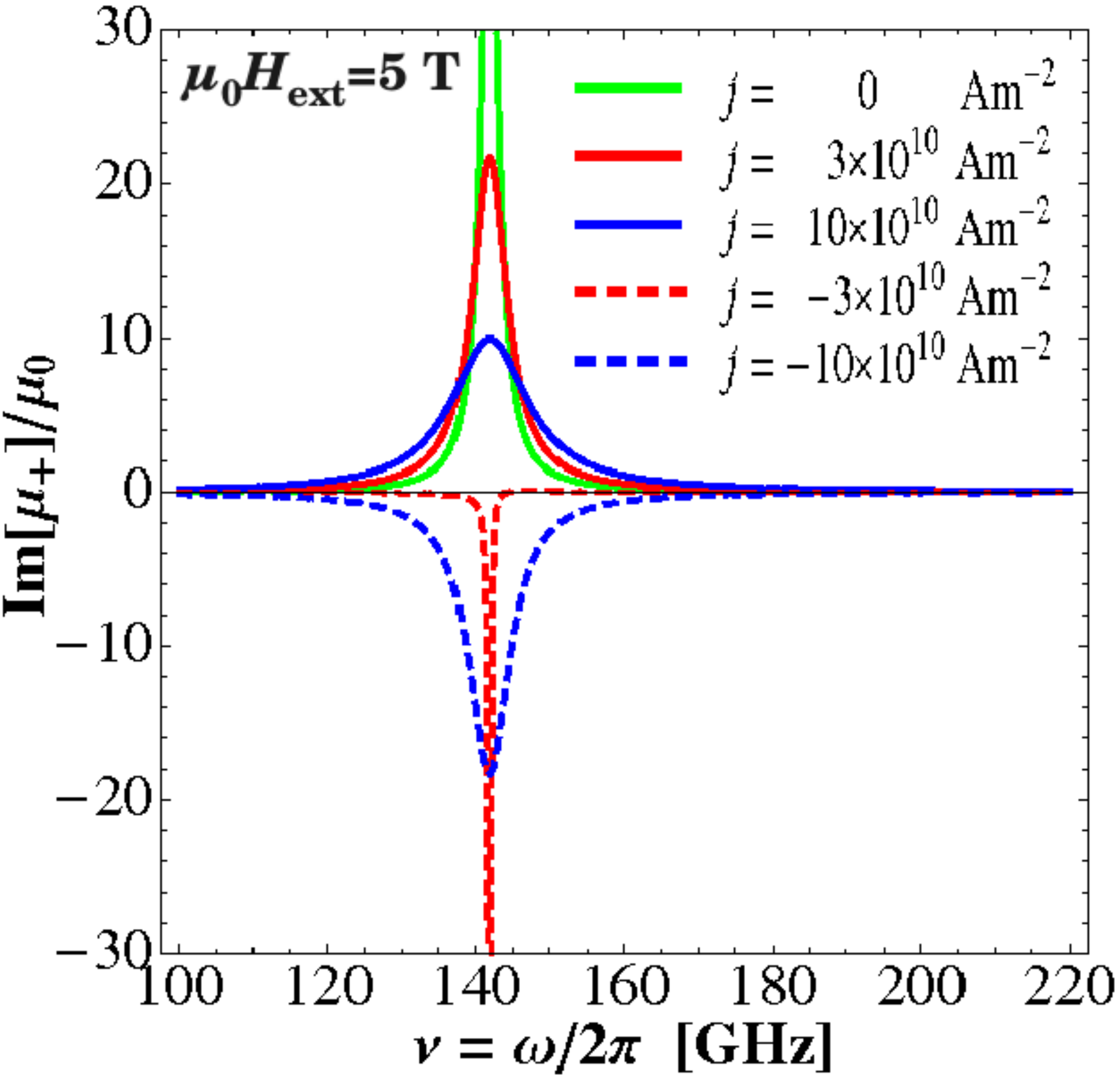}
\end{minipage}
\caption{ 
Real and imaginary parts of magnetic permeability for left-handed wave, $\mu_+$, as function of frequency, $\nu\equiv\omega/(2\pi)$, at $\muz\Hext=5$T ($\hext=\muz\gamma\Hext=880$GHz). $\omegaM=\muz\frac{\hbar\gamma^2 S}{a^3}$ is 388GHz.
The imaginary part becomes negative for a negative current satisfying \Eqref{imzero}.
\label{FIGReImomega}}
\end{figure}

We consider a circularly polarized incident wave. 
The magnetic field $\hvem$ then is
$\hvem=h_0 \lt( 1 , \pm i \rt)$, 
where $h_0$ is the amplitude and the sign of $\pm$ represents left- and right-handed polarization (or positive and negative helicity), respectively \cite{Jackson98}.
Using $\sigma_y \hvem=\pm \hvem$, we obtain the real and imaginary parts of the permeability for each circular polarization, ${\mu_\pm }$, as
\begin{align}
\Re \mu_\pm  
&=  {\muz} \lt[ 1 + \omegaM 
  \frac{ \mp\omega+\hext + \beta \jtil }
    {(\mp\omega+\hext + \beta \jtil)^2+(\omega\alpha \pm  P\jtil)^2 }  \rt]\nnr
\Im \mu_\pm 
 &= \muz \omegaM 
 \frac{\omega\alpha \pm P\jtil}
    {(\mp\omega+\hext+\beta \jtil)^2+(\omega\alpha \pm P\jtil)^2 } .
\label{ReImmu}
\end{align}
As seen from Eq. (\ref{muresult1}) and Eq. (\ref{ReImmu}), the resonance arises only for a left-handed light when $\Hvext$ is along positive $z$ direction.
Below, we consider only the left-handed case, $\mu_+$;
\begin{align}
\Re \mu_+  
&=  {\muz} \lt[ 1 + \omegaM 
  \frac{ -\omega+\hext + \beta \jtil }
    {(-\omega+\hext + \beta \jtil)^2+(\omega\alpha + P\jtil)^2 }  \rt]\nnr
\Im \mu_+ 
 &= \muz \omegaM 
 \frac{\omega\alpha + P\jtil}
    {(-\omega+\hext + \beta \jtil)^2+(\omega\alpha + P\jtil)^2 } .
\label{ReImmu_left}
\end{align}
Numerical result is shown in Fig. \ref{FIGReImomega}.
Parameters are chosen as $S=1$, $a=2.2$\AA, resulting in 
$\muz M=\muz\frac{\hbar \gamma}{a^3}S=2.2$T and $\omegaM=388$GHz, and $P=1$ and $\alpha=\beta=0.01$.

Let us first look into the imaginary part. 
We see that when current is zero, it is positive definite, indicating that a damping of spin (represented by $\alpha$) results in a loss of electromagnetic wave.
This is no longer true when current is switched on;
if $P\jtil < -\omega\alpha$, the imaginary part becomes negative for a left-handed wave.
The incident wave is therefore amplified by coupling to the current-driven magnetization, if the gain overcomes the electric loss described by the imaginary part of the permittivity.
Remarkably, amplification occurs for a particular polarization, either left or right, and is controlled by the current direction, as seen from Eq. (\ref{ReImmu}).
The present spin-torque oscillator system is thus an active filter to obtain a particular circular polarization.
The energy necessary for amplification is supplied from the applied current.

Let us look in detail why spin-transfer torque in the present system results in a gain.
When the fixed magnetization and external field is along $z$ direction, the field-induced torque $\torque_{\rm H}$ and current-induced field-like torque $\torque_{\rm F}$ acting on $\Mv$ are along a tangent to precession motion (Fig. \ref{FIGmeta_system}(b)).
The Gilbert damping torque is perpendicular to these torques and tends to suppress the precession amplitude, pointing $\Mv$ eventually in the equilibrium direction along $z$ axis.
The spin-transfer torque in the present configuration is parallel or antiparallel to the damping torque depending on the sign of the current.
When spin-transfer torque points opposite to the damping torque, negative damping is realized resulting in a gain.

Let us turn to the real part of Eq. (\ref{ReImmu_left}).
As a function of $\omega$, the real part of $\mu_{+}$ for a left-handed case has a maximum or minimum 
of (using $\alpha \ll 1$)
\begin{align}
\Re  \mu_+(  \omega_\mp )
& \simeq 
\muz \lt(1 \pm   \frac{\omegaM}{2P\jtil} \rt).
\end{align}
at
$\omega \simeq
\hext+\beta \jtil \mp P\jtil \equiv \omega_{\mp}$.
The real part therefore becomes negative around $\omega=\omega_+=\hext+\beta \jtil + P\jtil$ if $0<P\jtil< \frac{\omegaM}{2}$ and around 
$\omega=\omega_-=\hext+\beta \jtil + |P\jtil|$ if 
$ -\frac{\omegaM}{2}< P\jtil< 0$ 
(see Fig. \ref{FIGReImomega}(a)).

\begin{figure}\centering
\begin{minipage}{0.45\hsize}
\includegraphics[width=0.95\hsize]{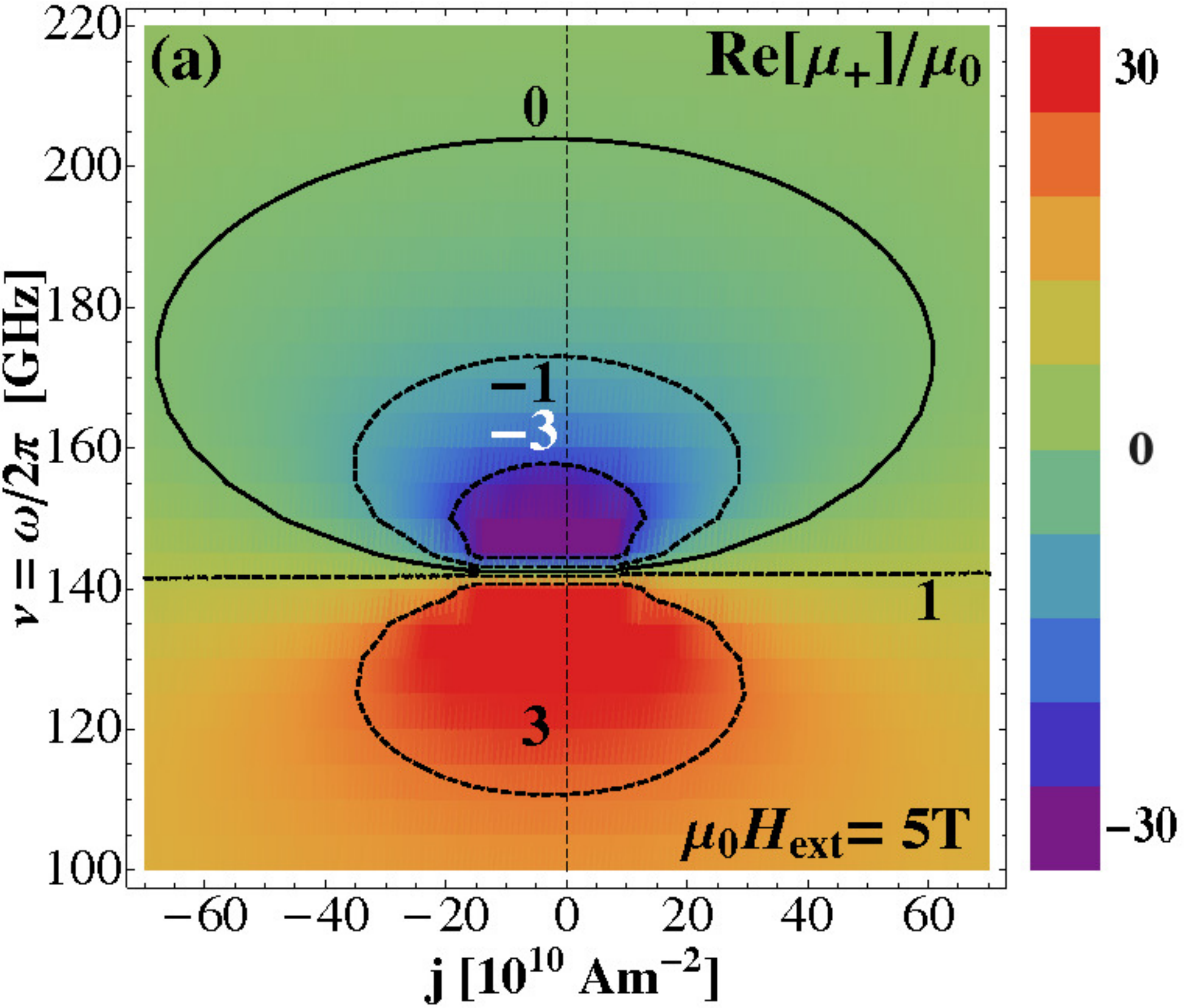}
\end{minipage}
\begin{minipage}{0.45\hsize}
\includegraphics[width=0.95\hsize]{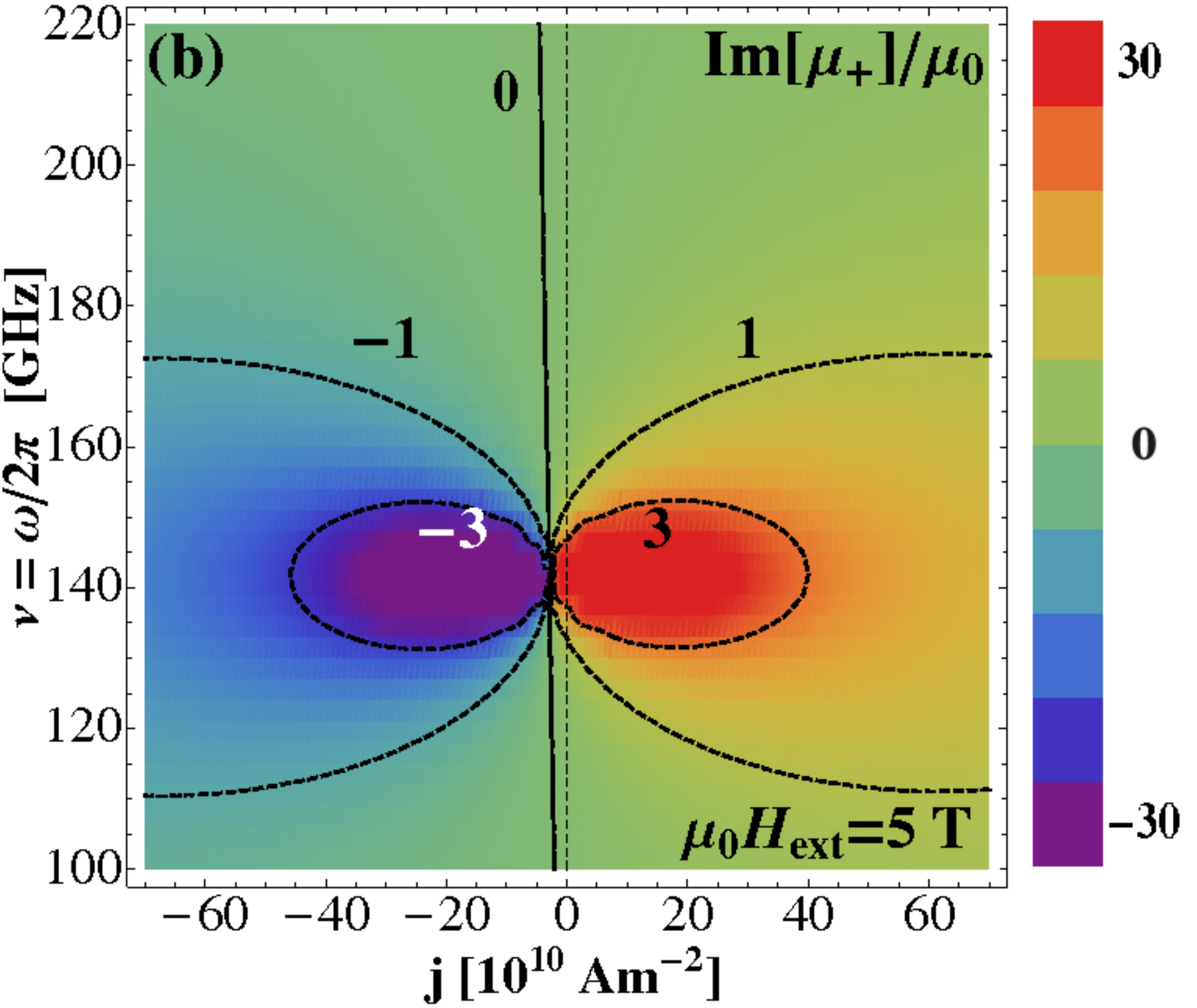}
\end{minipage}
\begin{minipage}{0.5\hsize}
\includegraphics[width=0.95\hsize]{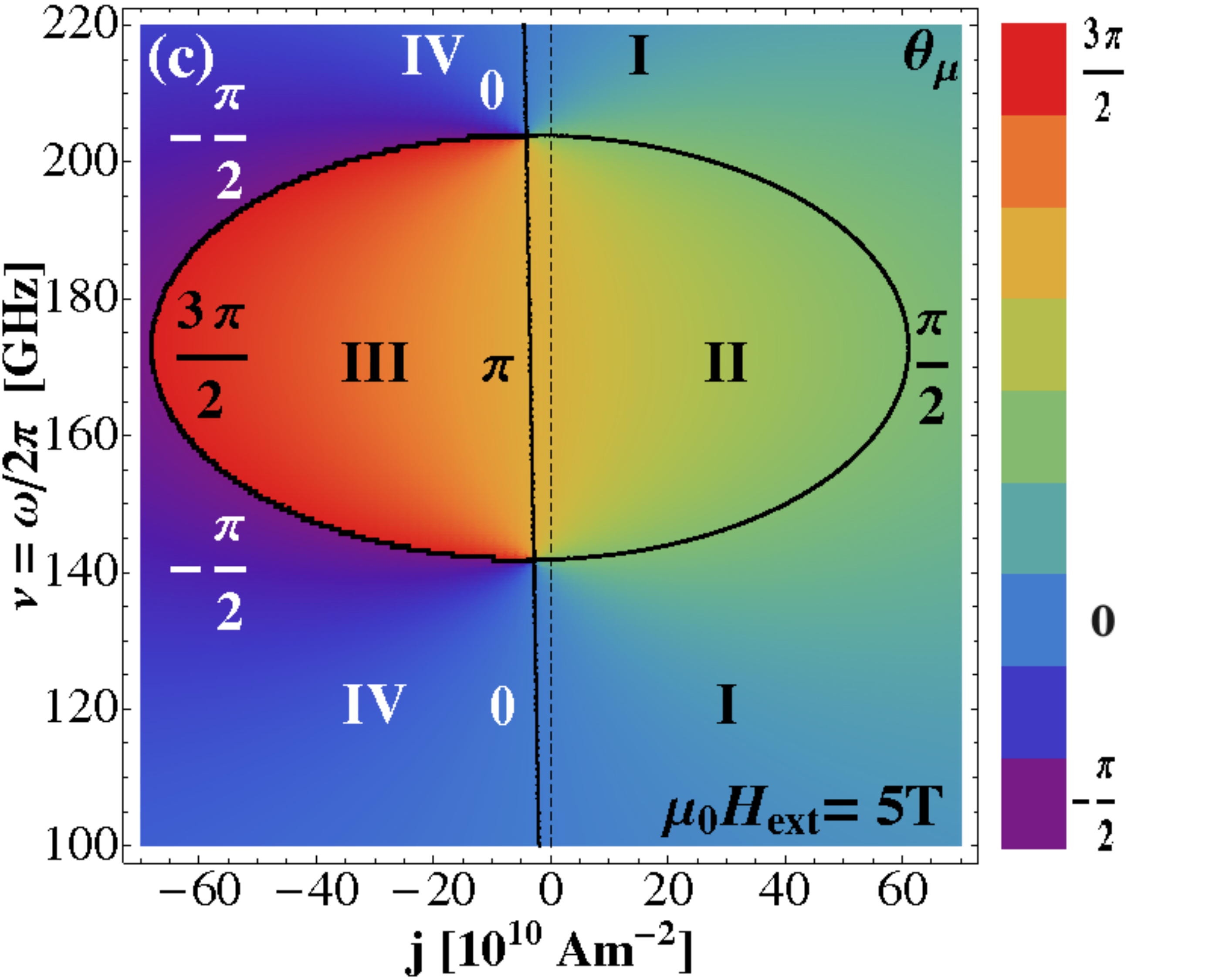}
\end{minipage}
\caption{ 
Contour plots of (a) real and (b) imaginary parts of $\mu_+$ and (c) phase $\theta_\mu$ in a $\nu-j$ plane at $\muz\Hext=5$T.
Regions I-IV in (c) are defined as follows. I): $\Re \mu_+>0$ and $\Im \mu_+>0$, II): $\Re \mu_+ <0$, $\Im \mu_+>0$, 
III): $\Re \mu_+ <0$, $\Im \mu_+<0$ and IV): $\Re \mu_+ >0$, $\Im \mu_+<0$.
The phase is defined in the regime 
$-\frac{\pi}{2}+\delta_\epsilon <\theta_\mu < \frac{3\pi}{2}+\delta_\epsilon$ 
(see Eq. (\ref{thetamucond})), where 
$\delta_\epsilon$ is a small quantity defined by \Eqref{deltaepsilondef}.
The phase has a discontinuity between regions III and IV.
\label{FIGomega-j}}
\end{figure}

Let us look in detail the behavior of $\mu_+$ as function of applied current and frequency.
Equation (\ref{ReImmu_left}) indicates that negative imaginary part arises when 
\begin{align}
\jtil < -\frac{\alpha}{P}\omega , \label{imzero}
\end{align}
 as depicted by a straight line in Fig. \ref{FIGomega-j}(b).
It is seen that negative imaginary part, i.e., possible gain material, is realized for a broad region of 
$\jtil >0$. 
This is because the original damping constant $\alpha$ is usually small (typically $\alpha\simeq 0.01$).
One should note, however, that the absolute value of imaginary part is small away from the resonance (Figs. \ref{FIGReImomega} and \ref{FIGomega-j}).

As for the real part of $\mu_+$,  we see from Eq. (\ref{ReImmu_left}) that it vanishes when 
\begin{align}
\lt[\omega-\lt(\frac{\omegaM}{2}+\hext\rt)\rt]^2+(P\jtil)^2
=\lt(\frac{\omegaM}{2}\rt)^2 .
\label{rezero}
\end{align}
Here we assumed $|\beta \jtil| \ll \omegaM$ and $\alpha \omega \ll |P\jtil|$.
Negative $\Re \mu_+$ is thus realized inside a oval shown in Fig.  \ref{FIGomega-j}(c).
From Eqs. (\ref{imzero}) and (\ref{rezero}), we see that there are four distinct regions, namely,
I) $\Re \mu_+>0$ and $\Im \mu_+>0$, II) $\Re \mu_+ <0$, $\Im \mu_+>0$, 
III) $\Re \mu_+ <0$, $\Im \mu_+<0$ and IV) $\Re \mu_+ >0$, $\Im \mu_+<0$ (Fig. \ref{FIGomega-j}(c)).

\section{Permittibity \label{SEC:epsilon}}

So far we discussed solely the magnetic property.
Let us now discuss the refractive index, $n\equiv \sqrt{\mu\epsilon/\muz \ez}$, by including the property of permittivity, $\epsilon$.
In the microwave regime, $\epsilon$ in metals has a large imaginary part due to a strong dissipation and negative real part, reflecting the fact that the electromagnetic wave is evanescent.
This fact is expressed in equations as follows. 
The incident electric field $\Ev$ induces a charge current density $\jv=\sigma(\omega)\Ev$, where $\sigma(\omega)$ is conductivity, resulting in a decay of $\Ev$.
Combining the Ohm's law with two of the Maxwell's equations, $\nabla\times\Ev=-\frac{\partial \Bv}{\partial t}$ and $\nabla\times\Bv=\muz\jv+\muz\ez \frac{\partial \Ev}{\partial t}$, 
we obtain the permittivity modified by the Ohm's law as (using $\omega\tau\ll1$, where $\tau$ is the elastic lifetime of electron)
\begin{align}
\epsilon=\ez\lt(1+\frac{i\sigma(\omega)}{\ez\omega}\rt),
\end{align}
neglecting here the cyclotron motion due to the magnetic field.
The imaginary part of $\epsilon$ is positive, resulting in a finite penetration length of $l=(\Im \sqrt{\epsilon\muz}\omega)^{-1}$ and 
zero transmission of propagating wave \cite{Jackson98}.
The conductivity in the microwave regime is described by the Drude model assuming free electrons, 
\begin{align}
\sigma(\omega)=\sigmaB \frac{1}{1-i\omega\tau}, 
\end{align}
where $\sigmaB\equiv \frac{e^2 n \tau}{m}=\ez\omegaP^2\tau$ is the Boltzmann conductivity and 
 $\omegaP\equiv\sqrt{\frac{e^2 n}{\ez m}}$ is the plasma frequency
\cite{Jackson98}.
Including the effect of cyclotron motion, the permittivity is modified to be polarization-dependent as \cite{Veselago68} (see Appendix \ref{SEC:app_epsilon})
\begin{align}
\epsilon_\pm\equiv \ez\lt(1-\frac{\omegaP^2}{\omega}\frac{1}{\omega\mp\omegac+\frac{i}{\tau}}\rt),
\end{align}
where $\omegac\equiv \frac{eB}{m}=\hext+\omegaM$ is the cyclotron angular frequency (electron charge is $-e<0$).
In the low frequency case we consider, $\omega,\omegac \ll\frac{1}{\tau}$, 
we thus obtain
\begin{align}
\Re \epsilon_\pm\simeq\ez\lt(1-(\omegaP\tau)^2\frac{\omega\mp\omegac}{\omega}\rt),
\end{align}
 and 
\begin{align}
\Im \epsilon_\pm\simeq\frac{\sigmaB}{\omega}=\ez\frac{\omegaP^2\tau}{\omega}.
\end{align}
Therefore, the phase $\theta_{\epsilon_\pm}$, defined by  $\epsilon_\pm=|\epsilon_\pm|e^{i\theta_{\epsilon_\pm}}$, is 
\begin{align}
\theta_{\epsilon_\pm}=\tan^{-1}\frac{\omegaP^2 \tau}{\omega[1-(\omegaP\tau)^2\frac{\omega\mp\omegac}{\omega}]}.
\end{align}
Due to strong dissipation by eddy current, $\theta_\epsilon$ in the GHz range is close to $\frac{\pi}{2}$.
Choosing $\kf^{-1}=0.8$\AA, we have  $\ef=5.5$eV and $\omegaP=1.4\times10^{16}$Hz ($\hbar\omegaP=9.0$eV).
Assuming dirty metals, we choose $\ef\tau/\hbar=10$, resulting in 
$\omegaP\tau =1.6$.
For $\muz\Hext=5$T and $\nu=\omega/(2\pi)=200$GHz, we obtain  
$\epsilon_+ /\ez\simeq 1.0 +i\times1.8\times10^4$.
Since permittivity has large positive imaginary part, we define the phase for right-handed polarization, $\theta_{\epsilon_+}$, as
\begin{align}
\theta_{\epsilon_+}=\frac{\pi}{2}+\delta_\epsilon.\label{deltaepsilondef}
\end{align}
The magnitude of deviation, $|\delta_\epsilon|$ is then usually very small; typically of the order of $10^{-4}$.

As seen from Fig. \ref{FIGomega-j}(c) and \Eqref{rezero}, interesting possibility of negative real part of $\mu_+$ arises in the region $\omega<\omegaM+\hext=\Omega_c$.
Therefore, in this region, the real part of $\epsilon_+$ is positive and hence $\delta_{\epsilon}<0$.
If the carrier is hole with positive charge, sign of $\Omega_c$ reverses, and 
$\Re \epsilon_+<0$  and $\delta_\epsilon>0$ are realized (if $\omegaP\tau\gtrsim1$).
As we will show in the next section, the sign of $\delta_\epsilon$ is important; 
it determines whether the system is normal (real part of refractive index is positive) or meta (negative real part) in the active region.


\section{Refractive index \label{SEC:n}}

\subsection{Determination of phase of refractive index}

The Maxwell's equations impose a relation between the wave vector and angular frequency as
$k^2=\epsilon\mu\omega^2$, where $\mu$ and $\epsilon$ are permeability and permittivity, respectively.
The relative refractive index is thus given by 
\begin{align}
n=\pm\lt(\frac{\epsilon\mu}{\ez\muz}\rt)^{\frac{1}{2}}.
\end{align}
The sign here, and more precisely the phase of $n$ in the complex plane is crucially important in discussing the active nature and sign of $\Re n$.
We here demonstrate that the phase is determined by the boundary condition imposing that the energy flow is continuous at the interface.
We consider an interface perpendicular to $z$ axis.
Two of the Maxwell's equations, 
$\nabla\times\Ev=-\mu\frac{\partial \Hv}{\partial t}$
and $\nabla\times \Hv=\epsilon \frac{\partial\Ev}{\partial t}$
then leads to the continuity of $E_x,E_y$, $H_x$ and $H_y$.
We consider for simplicity an incident radiation perpendicular to the interface.
The first of the above two equation reads in terms of wave vector, $\kv=(0,0,k)$, as
$\kv\times\Ev=\omega\mu\Hv$, i.e.,
$(H_x,H_y)=\frac{n}{\mu}(-E_y,E_x)$, where $n\equiv \frac{k}{\omega}$.
The energy flow represented by the Poynting vector along $z$ axis is
$P_z=(\Ev\times\Hv)_z=\frac{n}{\mu}(E_x^2+E_y^2)$.
The sign of the real part of $n/\mu$ thus determines the direction of the energy flow, and it needs to be positive to describe the case where the incident radiation enters the material.
In terms of the phase, defined as  $\epsilon=|\epsilon|e^{i\theta_\epsilon}$, $\mu=|\mu|e^{i\theta_\mu}$ and $n=|n|e^{i\theta_n}$, this boundary condition requires that
\begin{align}
-\frac{\pi}{2}<\theta_n-\theta_\mu <\frac{\pi}{2}.
\end{align}

The case of $|\theta_n-\theta_\mu|=\frac{\pi}{2}$ corresponds to a perfect reflection, since $\Re P_z$ vanishes.
Since $\theta_n=\frac{1}{2}(\theta_\mu+\theta_\epsilon)$, the above condition reads
\begin{align}
-{\pi}<\theta_\mu-\theta_\epsilon <{\pi}.
\end{align}

In the present metallic system, $\theta_\epsilon$ is close to $\frac{\pi}{2}$ in the microwave regime, and thus
$\theta_\epsilon=\frac{\pi}{2}+\delta_\epsilon$, where $\delta_\epsilon$ is a small deviation (positive or negative) as explained in Sec. \ref{SEC:epsilon}.
The phase $\theta_\mu$ is then lies in the region (Fig. \ref{FIGomega-j}(c))
\begin{align}
-\frac{\pi}{2}+\delta_\epsilon <\theta_\mu < \frac{3\pi}{2}+\delta_\epsilon,
\label{thetamucond}
\end{align}
 and $\theta_n$ takes a value of (Fig. \ref{FIGphase_n}(a))
\begin{align}
\delta_\epsilon <\theta_n < \pi +\delta_\epsilon.\label{thetancond}
\end{align}

From Eq. (\ref{thetancond}), when $\delta_\epsilon>0$, we have three possibilities; 
1) normal dissipative, corresponding to $\delta_\epsilon <\theta_n < \frac{\pi}{2}$
( $-\frac{\pi}{2}+\delta_\epsilon <\theta_\mu < \frac{\pi}{2}-\delta_\epsilon$),
2) meta dissipative, corresponding to $\frac{\pi}{2} <\theta_n < \pi $
( $\frac{\pi}{2}-\delta_\epsilon <\theta_\mu < \frac{3\pi}{2}-\delta_\epsilon$),
and
3) meta gain, corresponding to $ \pi <\theta_n < \pi +\delta_\epsilon$
( $\frac{3\pi}{2}-\delta_\epsilon <\theta_\mu < \frac{3\pi}{2}+\delta_\epsilon$).
Only the third case is the active case, but the window for this region is very narrow.
When $\delta_\epsilon<0$, we have 
1') normal dissipative, corresponding to $0 <\theta_n < \frac{\pi}{2}$
( $-\frac{\pi}{2}-\delta_\epsilon <\theta_\mu < \frac{\pi}{2}-\delta_\epsilon$),
2') meta dissipative, corresponding to $\frac{\pi}{2} <\theta_n < \pi -|\delta_\epsilon|$
( $\frac{\pi}{2}-\delta_\epsilon <\theta_\mu < \frac{3\pi}{2}+\delta_\epsilon$),
and
3') normal gain, corresponding to $ -|\delta_\epsilon| <\theta_n < 0$
( $-\frac{\pi}{2}+\delta_\epsilon <\theta_\mu < - \frac{\pi}{2}-\delta_\epsilon$).
Again the active case arises in a very narrow window with width of $\Delta \theta_\mu\simeq 2\delta_\epsilon$.

From the above argument, we see that spin-torque oscillator system realizes negative refraction in rather a broad area in the $\omega$-$j$ plane, while active material in contrast needs fine tuning of applied current to satisfy 
$ \frac{3\pi}{2} -\delta_\epsilon< \theta_\mu < \frac{3\pi}{2}+\delta_\epsilon$ (if $\delta_\epsilon>0$)
or $ -\frac{\pi}{2}-|\delta_\epsilon| < \theta_\mu < - \frac{\pi}{2}+|\delta_\epsilon|$ (if $\delta_\epsilon<0$).
As we have seen in the preceding section, $\delta_\epsilon<0$ in the case of electron-like carrier, while $\delta_\epsilon>0$ in the hole-like case, and 
thus active regime is normal (positive $\Re n$) for electron while it is meta ($\Re n<0$) for hole.

\subsection{Refractive index of spin-torque oscillator}

\begin{figure}\centering
\includegraphics[width=0.4\hsize]{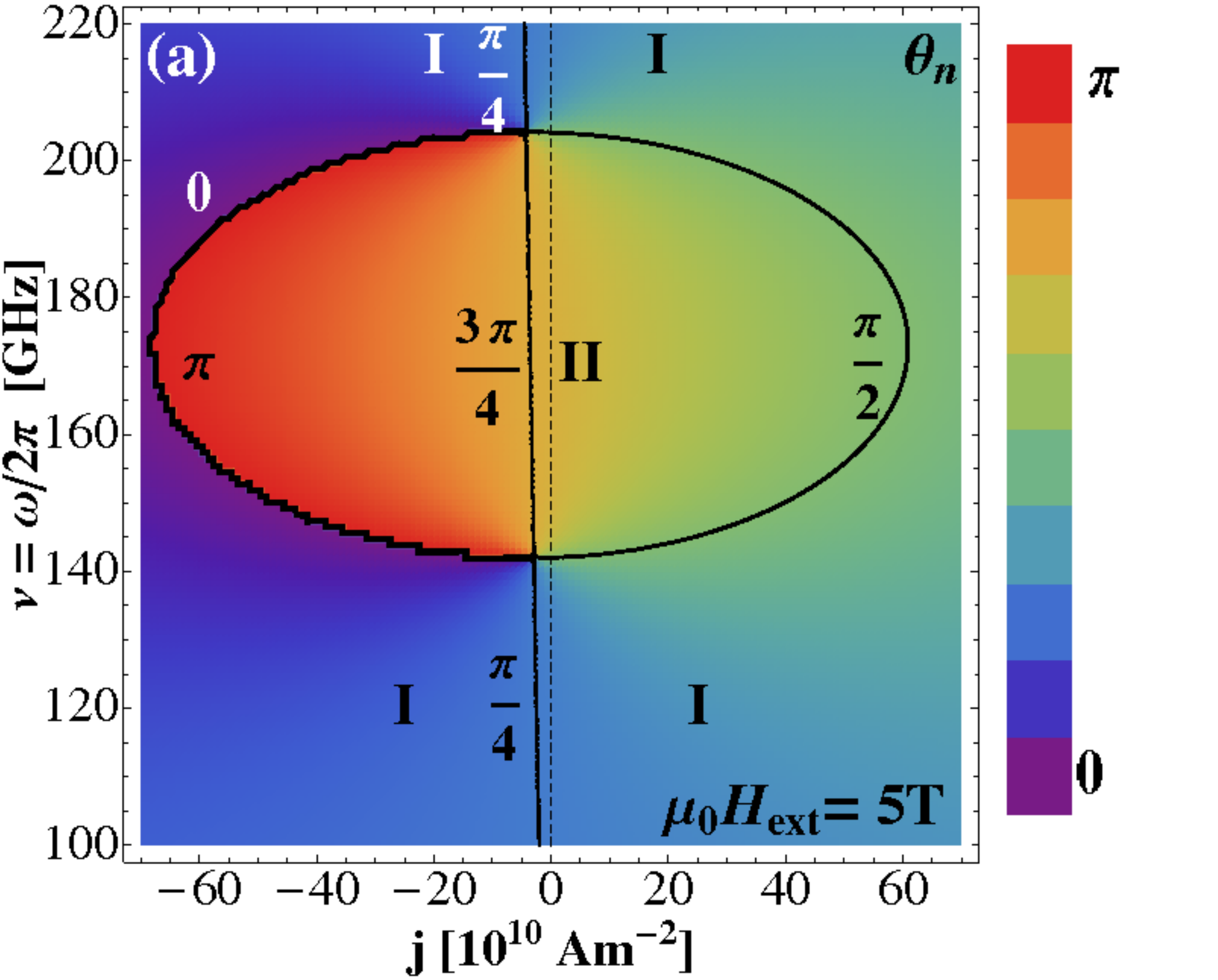}
\includegraphics[width=0.4\hsize]{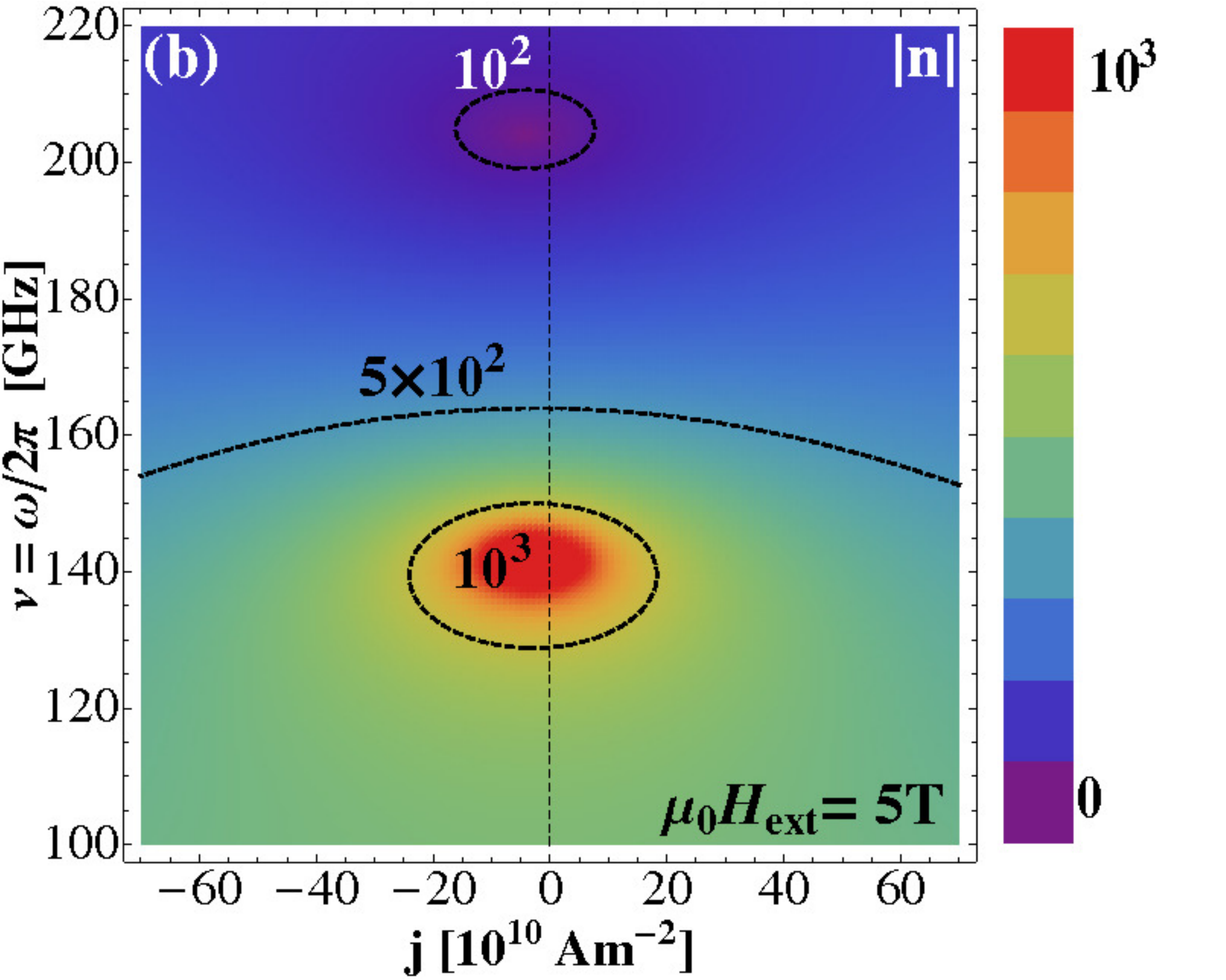}
\includegraphics[width=0.4\hsize]{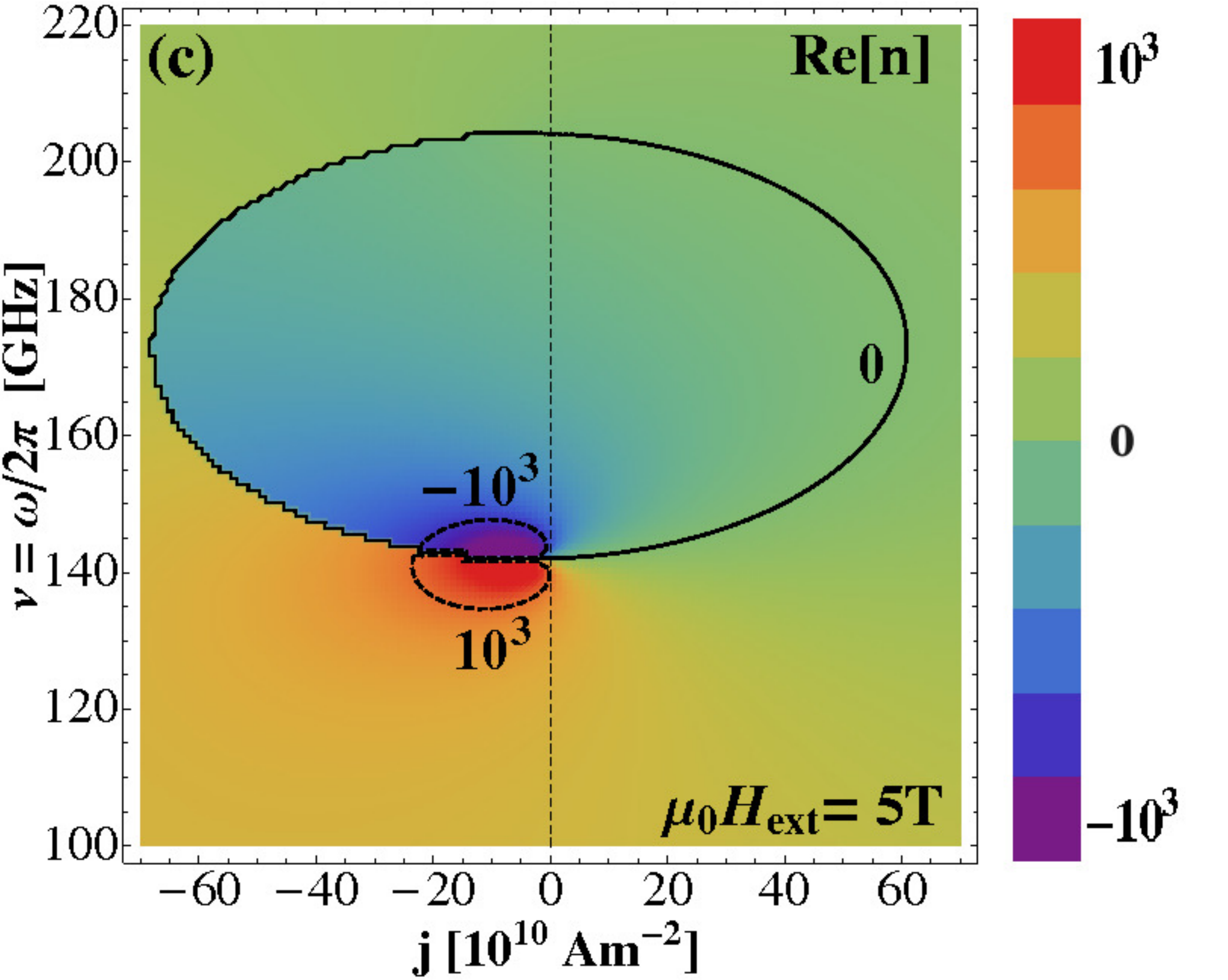}
\includegraphics[width=0.4\hsize]{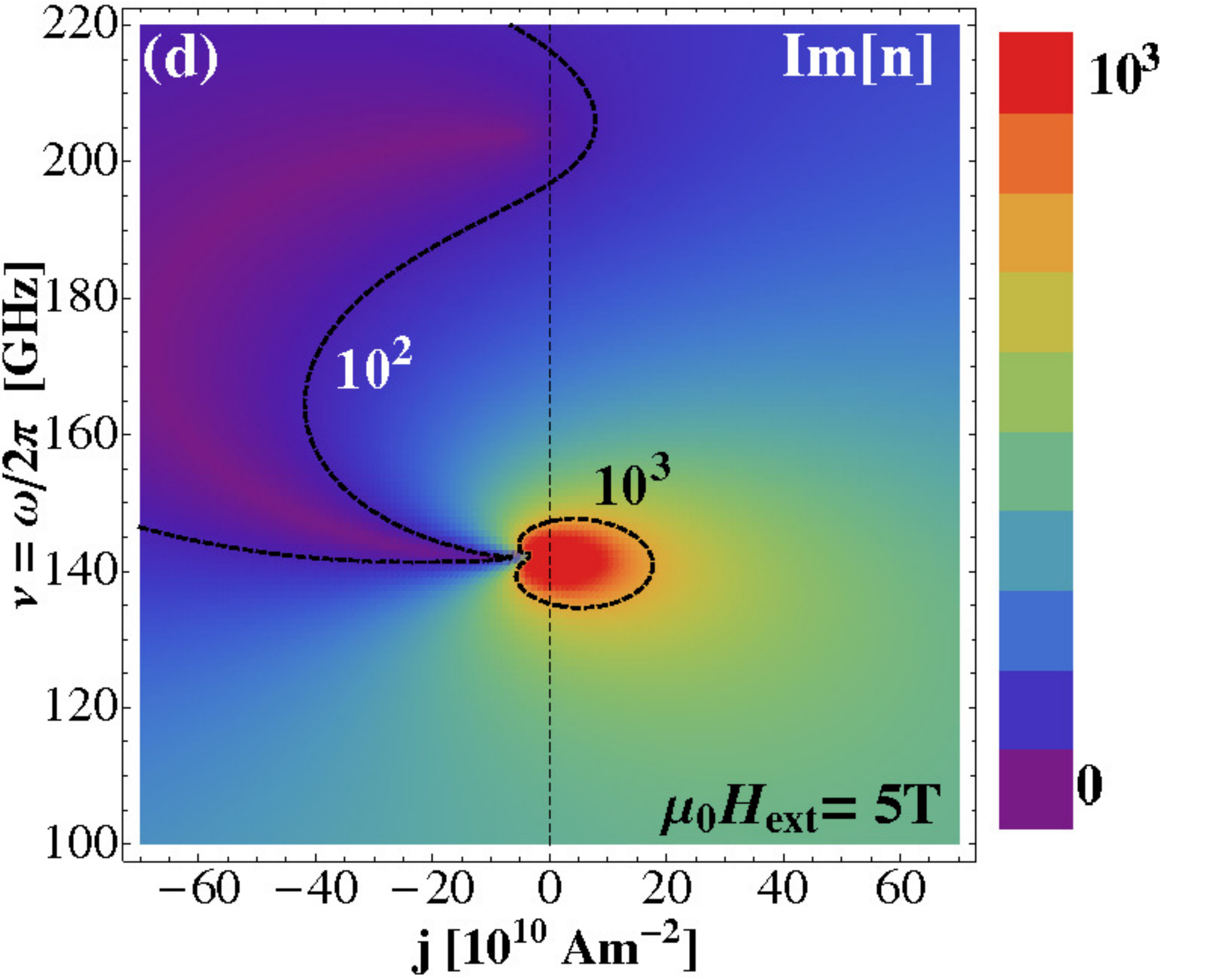}
\caption{ 
(a) Phase, (b) absolute value, (c) real part and (d) imaginary part of refractive index for an incident light with left-handed polarization in a spin-torque oscillator device in the $\nu$-$j$ plane. 
Region I and II are $\Re \mu_+>0$ and $\Im \mu_+>0$, and $\Re \mu_+ <0$ and $\Im \mu_+>0$, respectively.
Phase is defined in the regime $\delta_\epsilon <\theta_n<\pi+\delta_\epsilon$ (\Eqref{thetancond}), and there is 
a discontinuity of phase from $\delta_\epsilon$ to $\pi+\delta_\epsilon$ (represented by a  thick line in Fig. \ref{FIGphase_n}(a)). 
Active region is realized very close to the discontinuity for the region satisfying \Eqref{activecondition}.  
(The active region is very narrow and not recognizable in the figure).
\label{FIGphase_n}}
\end{figure}

\paragraph{Dissipative region}

Figure \ref{FIGphase_n} shows results of refractive index, $n$, of spin-torque oscillator.
The magnitude of $n$ (Fig. \ref{FIGphase_n}(b)) is generally large because of large imaginary part of permeability in the GHz range.
Negative real part, i.e., electromagnetic metamaterial, is realized in the regime inside the oval determined by \Eqref{rezero} (Fig. \ref{FIGphase_n}(c)).
Most area outside of the oval is regime 1) or 1') in the preceding subsection, namely, it is normal dissipative (ND) regime.
Inside the oval, meta dissipative (MD) is realized corresponding to regime 2) or 2'). 
As we have discussed in the preceding subsection, there is a discontinuity of the phase $\theta_\mu$ 
 at the boundary of the oval in the regime $\jtil<-\alpha \omega/P$ from $\frac{3\pi}{2}$ to $-\frac{\pi}{2}$, and 
active material (negative imaginary part) is realized in a very narrow window around the discontinuity represented by a thick line in Fig. \ref{FIGphase_n}(a) (the active regime is not recognizable in the figure).

Appearance of negative refractive index (electromagnetic meta material) 
in a broad region around $j=0$ in Fig. \ref{FIGphase_n}(c) is consistent with recent experimental discovery of  negative refractive index in Co \cite{Engelbrecht11}.

\paragraph{Active (gain) region}

Let us study the active regime (case 3) or 3')) in detail.
The condition for the active regime is 
\begin{align}
|\delta_\mu|\lesssim|\delta_\epsilon|,
\end{align}
where $\delta_\mu\equiv \theta_{\mu_+}+\frac{\pi}{2}$.
Since $\Re \mu_+ \sim0$ in this regime, $\delta_\mu$ is given by
\begin{align}
\delta_\mu \simeq \frac{\Re \mu_+}{\Im \mu_+}.
\end{align}
From \Eqref{rezero}, the current density realizing $\delta_\mu=0$ ($\Re \mu_+ =0$) for $\jtil<0$ is given by 
$\jtil=\jtil_{\rm c}\equiv 
-\frac{1}{P}[\alpha\omega+\sqrt{(\omega-\hext)(\omegac-\omega)}]$.
Defining small deviation from $\jtil_{c}$ as 
$\jtil=\jtil_{c}+\delta\jtil$, 
The real and imaginary parts read
\begin{align}
{\Re \mu_+}/\muz & 
 \simeq \frac{2P(P\jtil_{c}+\alpha\omega)}{\omegaM(\omega-\hext)} \delta \jtil 
  \nnr
{\Im \mu_+}/\muz & 
 \simeq \frac{P\jtil_{c}+\alpha\omega}{\omega-\hext}
    + \frac{4P}{\omegaM} \frac{\omega-\lt(\hext+\frac{\omegaM}{2}\rt)}{\omega-\hext}\delta \jtil ,
\end{align}
neglecting the higher order of $\delta\jtil$.
Thus 
\begin{align}
\delta_\mu \simeq \frac{2P}{\omegaM}\delta\jtil.
\end{align}
The condition for the active media, $|\delta_\mu|\lesssim|\delta_\epsilon|$, then becomes 
\begin{align}
|\delta j| \lesssim \frac{eS^2}{2Pa^2}\omegaM|\delta_\epsilon|.\label{activecondition}
\end{align}
Using the parameters used in Sec. \ref{SEC:epsilon},
$\frac{eS^2}{2Pa^2}\omegaM=6.4\times10^{11}$A/m$^2$.
The window for the active regime is thus $|\delta j| \lesssim 6\times 10^{7}$A/m$^2$ if we use $|\delta_\epsilon|=10^{-4}$.
Since the applied current density is of the order of $10^{11}$A/m$^2$ (see Fig. \ref{FIGphase_n}), therefore, fine tuning of current density within 0.1\% is necessary to realize active regime.
The active regime is normal (positive $\Re n$) for electron-like carrier while it is meta ($\Re n<0$) for the hole-like case.

The high accuracy required for active media is due to strong dissipation in the electric response ($\epsilon$) as we have seen. 
In reality, this condition is relaxed by several reasons.
Most obvious one is the lowering of the plasma frequency in dilute systems. In fact, spin-torque oscillators are usually made of small thin films with a diameter of 100nm or less \cite{Kiselev03}. 
For sufficient electromagnetic response, we need many of such small oscillators put on a substrate, resulting in the reduction of the effective electron density in the whole system.
The plasma frequency, proportional to the square root of the density, is therefore reduced. Reducing the effective density by a factor of 0.01 results in enhancement of $|\delta_\epsilon|$ by a factor of 10, and the accuracy needed is relaxed to be 1\%.
The effective mass of the electric oscillation can be tuned, too \cite{Pendry96}, and this helps the enhancement of $|\delta_\epsilon|$.

For an experimental realization of the present active material, we need to take into account the fact that spin-torque oscillators have thin free layers, with thickness less than 100nm, to excite a coherent oscillation of the magnetization.
A single free layer is therefore thinner than the penetration depth, which is of the order of micron meter, resulting in almost perfect transmission even in the dissipative regime.
To observe the active effect we have predicted, multilayer structures of spin-torque oscillators where the sum of the free layer is close to $\mu$m scale is necessary.
Besides, to place many spin-torque multilayers as an array is suitable for a large signal and also from the viewpoint of reducing dissipation, as we mentioned above.

\section{Summary}

We have shown that spin-torque oscillators act as an active media and/or an
electromagnetic metamaterial with negative refractive index for circularly
polarized microwaves.
Without spin-torque oscillators, metals thicker than the penetration depth 
 are almost perfect reflectors of microwaves. 
What we have shown above is that when the amplification by the current-driven magnetization precession in multilayers exceeds the dissipation due to the Ohm's law, current-driven ferromagnetic metal becomes transparent in the microwave regime. 
Experimental verification of this current-driven microwave switching is of great interest.

For experimental studies, films with a strong perpendicular easy axis anisotropy like FePt and CoPt are of particular interest, since the effective magnetic field ($\muz\Hext$) due to the anisotropy exceeds 10 T in these materials, and thus the present effects are realized at higher frequencies.

\section*{Acknowledgments}
G.T. thanks S. Tomita, T. Ueda, K. Sawada and M. Gonokami for valuable comments and discussions. 
This work was supported by a Grant-in-Aid for Scientific Research (B) (Grant No. 22340104) from Japan Society for the Promotion of Science and UK-Japanese Collaboration on Current-Driven Domain Wall Dynamics from JST. A.T. and K.T. are financially supported by the Japan Society for the Promotion of Science for Young Scientists.

\appendix

\section{Derivation of permittibity \label{SEC:app_epsilon} }

Equation of motion of the electron (with charge $q=-e$) under electric and magnetic fields is
\begin{align}
\dot{\bm{v}} & =  \frac{q\bm{E}}{m} + \frac{q}{m} \left( \bm{v} \times \bm{B} \right) - \frac{\bm{v}}{\tau}.
\end{align}
We consider the case $\Bv$ is along $z$ direction ($\Bv=(0,0,B)$ and $\Ev$ is in the $xy$ plane.
The equation then reads
\begin{align}
(1-i\omega\tau) v_x - \frac{q\tau}{m} v_y B = \frac{q\tau}{m} E_x, \\ 
(1-i\omega\tau) v_y + \frac{q\tau}{m} v_x B = \frac{q\tau}{m} E_y,
\end{align}
or
\begin{align}
	\left( 1-i\omega \tau - i\frac{q \tau B }{m} \sigma_y \right) \bm{v} = \frac{q \tau }{m} \bm{E}
\end{align}
For left- and right-haned polarization, $E_\pm\equiv E_0(1,\pm i)$, the velocity is
\begin{align}
	\bm{v}_{\pm}= \frac{ \frac{ q \tau }{m} }{ 1-i\omega \tau \mp i\frac{q \tau B }{m} } \bm{E},
\end{align}
or
\begin{align}
	\bm{v}_{\pm}= -\frac{ \frac{ e \tau }{m} }{ 1-i\omega \tau \pm i\frac{e \tau B }{m} } \bm{E}.
\end{align}
 
Using $\jv=-en\vv$ and $\Pv=\frac{i}{\omega}\jv$, $\epsilon=\ez+\frac{\Pv}{\Ev}$ is therefore obtained as
\begin{align}
	\epsilon_{\pm}= \ez \lt( 1 -\frac{ \omegaP^2 }{\omega} \frac{1}{ \omega \mp \Omega_c+ \frac{i}{\tau} } \rt) .
\end{align}


\bibliography{/home/tatara/References/12,/home/tatara/References/gt}

\end{document}